\documentclass[aps,prl,twocolumn,showpacs,preprintnumbers,amsmath,amssymb,superscriptaddress]{revtex4}%

\usepackage{graphicx}%
\usepackage{dcolumn}
\usepackage{amsmath}
\usepackage{color}

\begin{document}

\preprint{PREPRINT (\today)}

\title{Muon-spin rotation studies of SmFeAsO$_{0.85}$ and NdFeAsO$_{0.85}$ superconductors}

\author{Rustem~Khasanov}
 \email{rustem.khasanov@psi.ch}
 \affiliation{Laboratory for Muon Spin Spectroscopy, Paul Scherrer
Institut, CH-5232 Villigen PSI, Switzerland}
\author{Hubertus~Luetkens}
 \affiliation{Laboratory for Muon Spin Spectroscopy, Paul Scherrer
Institut, CH-5232 Villigen PSI, Switzerland}
\author{Alex~Amato}
 \affiliation{Laboratory for Muon Spin Spectroscopy, Paul Scherrer
Institut, CH-5232 Villigen PSI, Switzerland}
\author{Hans-Henning~Klauss}
 \affiliation{IFP, TU Dresden, D-01069 Dresden, Germany}
\author{Zhi-An~Ren}
 \affiliation{National Laboratory for Superconductivity, Institute of Physics and Beijing National Laboratory for Condensed Matter Physics, Chinese Academy of Sciences, P. O. Box 603, Beijing 100190, P. R. China}
\author{Jie Yang}
 \affiliation{National Laboratory for Superconductivity, Institute of Physics and Beijing National Laboratory for Condensed Matter Physics, Chinese Academy of Sciences, P. O. Box 603, Beijing 100190, P. R. China}
\author{Wei Lu}
 \affiliation{National Laboratory for Superconductivity, Institute of Physics and Beijing National Laboratory for Condensed Matter Physics, Chinese Academy of Sciences, P. O. Box 603, Beijing 100190, P. R. China}
\author{Zhong-Xian Zhao}
 \affiliation{National Laboratory for Superconductivity, Institute of Physics and Beijing National Laboratory for Condensed Matter Physics, Chinese Academy of Sciences, P. O. Box 603, Beijing 100190, P. R. China}

\begin{abstract}
Measurements of the in-plane magnetic field  penetration depth
$\lambda_{ab}$ in Fe-based superconductors with the nominal
composition SmFeAsO$_{0.85}$ ($T_c\simeq52$~K) and
NdFeAsO$_{0.85}$ ($T_c\simeq51$~K) were carried out by means of
muon-spin-rotation. The absolute values of $\lambda_{ab}$ at $T=0$
were found to be 189(5)~nm and 195(5)~nm for Sm and Nd substituted
samples, respectively. The analysis of the magnetic penetration
depth data within the Uemura classification scheme, which
considers the correlation between the superconducting transition
temperature $T_c$ and the effective Fermi temperature $T_F$,
reveal that  both families of Fe-based superconductors (with and
without fluorine) falls to the same class of unconventional
superconductors.
\end{abstract}
\maketitle


The recent discovery of  the Fe-based layered superconductor
LaO$_{1-x}$F$_x$FeAs \cite{Kamihara08} with the transition
temperature $T_c = 26$~K has triggered an intense research in the
oxypnictides. In its wake, a series of new superconductors with
$T_c$ onset of up to 55~K have been synthesized successively by
substituting La with other rare earth (Re) ions like Sm, Ce, Nd, Pr,
and Gd \cite{Ren08_SmF,Ren08_PrF}. Recently the new family
of the oxypnictide superconductors ReFeAsO$_{1-x}$ with the
doping induced by oxygen vacanies instead of fluorine substitution
were synthesized \cite{Ren08_F-free,Yang08}. The tunable oxygen content
which leads to the occurrence of superconductivity strongly
resembles the situation in high-temperature cuprate
superconductors (HTS's).

One of the interesting questions, which still awaits to be
explored, is to which class of the superconducting materials the
newly discovered Fe-based superconductors belong. The search for
relations between the various physical variables such as
transition temperature, magnetic field penetration depth,
electrical conductivity, energy gap, Fermi temperature {\it etc.}
may help to answer this question. Among others, there is a
correlation between $T_c$ and the zero-temperature inverse squared
magnetic field penetration depth [$\lambda^{-2}(0)$], that
generally relates to the zero-temperature superfluid density
($\rho_s$)  in terms of  $\rho_s\propto\lambda^{-2}(0)$. In
various families of underdoped HTS's there is the empirical
relation $T_c\propto\rho_s\propto\lambda^{-2}(0)$, first
identified by Uemura {\it et al.} \cite{Uemura89,Uemura91}. In
this respect it is rather remarkable that the first magnetic field
penetration measurements on LaO$_{1-x}$F$_x$FeAs ($x=0.1$, 0.075)
\cite{Luetkens08} and SmO$_{0.82}$F$_{0.18}$FeAs \cite{Drew08}
result in values of the superfluid density which are close to the
Uemura line for hole doped cuprates, indicating that the
superfluid is also very dilute in the oxypnictides.

In this paper we focus on the different classification  scheme
proposed by Uemura  \cite{Uemura91,Hillier97} which considers the
correlation between  $T_c$ and the effective Fermi temperature
$T_F$ determined from measurements of the in-plane magnetic
penetration depth $\lambda_{ab}$. Within this scheme strongly
correlated unconventional superconductors, as HTS's, heavy
fermions, Chevrel phases, or organic superconductors form a common
but distinct group, characterized by a universal scaling of $T_c$
with $T_F$ such that $1/10>T_c/T_F>1/100$. We show that within the
Uemura classification scheme both families of oxypnictide
superconductors (with and without fluorine) falls to the same
class of unconventional superconductors.

Details on the sample preparation for SmFeAsO$_{0.85}$ and NdFeAsO$_{0.85}$ can be
found elsewhere \cite{Ren08_F-free}.
The muon-spin rotation ($\mu$SR)  experiments were performed at
the $\pi$M3 beam  line at the Paul Scherrer Institute (Villigen,
Switzerland). During the experiments we were mostly concentrated
on SmFeAsO$_{0.85}$ which shows the highest $T_c$ among other
oxypnictide superconductors discovered till now. For
NdFeAsO$_{0.85}$ we studied only the temperature dependence of the
superfluid density in a field of 0.2~T.

\begin{figure}[htb]
\includegraphics[width=0.9\linewidth]{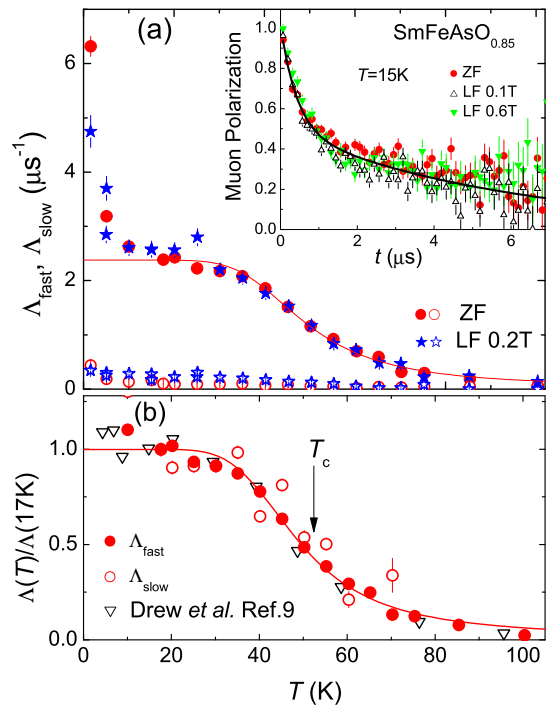}
\caption{(Color online) (a) Temperature dependences  of the fast
($\Lambda_{\rm fast}$) and the slow ($\Lambda_{\rm slow}$)
components of the ZF and the LF muon depolarization rate of
SmFeAsO$_{0.85}$. (b) $\Lambda_{\rm fast}(T)$ and $\Lambda_{\rm
slow}(T)$ normalized to their values at $T=17$~K. Open triangles
represents the normalized data of SmO$_{0.82}$F$_{0.18}$FeAs from
\cite{Drew08}.  The inset in (a) shows the ZF and the LF $\mu$SR
time-spectra of SmFeAsO$_{0.85}$ measured at $T=15$~K. The solid
lines in (a) and (b) are the fits of
Eq.~(\ref{eq:Activation-Energy}) to ZF $\Lambda_{\rm fast}(T)$.}
 \label{fig:ZF-LF}
\end{figure}

We first start our discussion with zero-field (ZF) and
longitudinal-field (LF) $\mu$SR experiments on SmFeAsO$_{0.85}$.
The recent ZF $\mu$SR studies of the parent LaOFeAs compound
reveal that there are two interstitial lattice sites where muons come into rest, namely, close to the Fe magnetic moments within the FeAs
layers and near the LaO layers \cite{Klauss08}. Bearing this in
mind the ZF and the LF muon-time spectra for $T\lesssim80$~K were
analyzed by using the following depolarization function:
\begin{equation}
P^{\rm ZF, LF}(t)=A_{\rm slow}\exp(-\Lambda_{\rm slow}t)+A_{\rm
fast}\exp(-\Lambda_{\rm fast}t).
 \label{eq:ZF-LF}
\end{equation}
Here $A_{\rm slow}$ ($A_{\rm fast}$) and $\Lambda_{\rm slow}$
($\Lambda_{\rm fast}$) are the asymmetry and the depolarization
rate of the slow (fast) component, respectively. The whole set of
ZF (LF) data was fitted simultaneously with the ratio $A_{\rm
slow}/A_{\rm fast}$ as a common parameter and the relaxations
($\Lambda_{\rm slow}$ and $\Lambda_{\rm fast}$) as individual
parameters for each particular data point. The total asymmetry
$A_{\rm slow}+A_{\rm fast}$ was kept constant within each set of
the data (ZF or LF). Above 80~K the fit becomes statistically
compatible with the single exponential component only. The results of
the analysis and the representative ZF and LF muon-time spectra
are shown in Fig.~\ref{fig:ZF-LF}.

From the data presented in Fig.~\ref{fig:ZF-LF} the following
important points emerge:
(i) Both $\Lambda_{\rm fast}(T)$ and  $\Lambda_{\rm slow}(T)$
measured in the zero and the longitudinal (up to 0.6~T) fields coincides
within the whole temperature region. This, together with the
exponential character of the muon polarization decay, reveal the
existence of fast electronic fluctuations measurable within the
$\mu$SR time-window. Assuming the fluctuation rate with a
temperature dependence $\nu\sim\exp(E_0/k_BT)$ ($E_0$ is the
activation energy) and accounting for the saturation of
$\Lambda\simeq\Lambda_0$ at 10~K$\lesssim T\lesssim$35~K the
relaxation is expected to follow \cite{Lancaster04}:
\begin{equation}
\frac{1}{\Lambda}=\frac{1}{\Lambda_0}+\frac{1}{C\exp(E_0/k_BT)}
 \label{eq:Activation-Energy}
\end{equation}
The fit of Eq.~(\ref{eq:Activation-Energy})  to the experimental
ZF $\Lambda_{\rm fast}$  data yields $\Lambda_{0,{\rm
fast}}=2.38(6)$~$\mu$s$^{-1}$, $C=0.012(2)$~$\mu$s, and
$E_0=23(2)$~meV. This kind of activated process can be anticipated for a thermal population of Sm crystal field levels.
(ii) The fact that both the slow and the fast relaxation rates
exhibit similar temperature dependences [see
Fig.~\ref{fig:ZF-LF}~(b)] strongly suggests that there is a common
source for both relaxations which most probably relates to the
fluctuation of Sm electronic moments. The magnitudes  of
$\Lambda_{\rm fast}$ and $\Lambda_{\rm slow}$ are thus related to
the different couplings between muons and Sm moments at the
distinct muon stopping sites.
(iii) There are no features appearing in the vicinity of the
superconducting  transition. Fig.~\ref{fig:ZF-LF}~(b) implies that
$\Lambda_{\rm fast}$ increases continuously with decreasing
temperature. This may suggest that the magnetic fluctuations
responsible for the effects seen in Fig.~\ref{fig:ZF-LF} are not
related to superconductivity. It is worth to mention that in
systems exhibiting an interplay between the superconductivity and
magnetism the slowing down of the spin fluctuations (increase of
$\Lambda$) correlates with $T_c$ (see e.g.
Ref.~\onlinecite{Amato03}). Another argument comes from the
comparison of $\Lambda_{\rm fast}(T)$ for SmFeAsO$_{0.85}$
($T_c\simeq52$~K), studied here, with that reported by Drew {\it
et al.} \cite{Drew08} for SmO$_{0.82}$F$_{0.18}$FeAs
($T_c\simeq45$~K), see Fig.~\ref{fig:ZF-LF} (b). Apparently the Sm
spin fluctuations are independent on $T_c$, which, therefore,
suggests that  their slowing down is not related to the
superconductivity.
(iv) The fast increase of both $\Lambda_{\rm fast}$ and
$\Lambda_{\rm slow}$ below 5~K is most probably associated with
additional local-field broadening due to the ordering of the Sm
moments \cite{Drew08,Ding08}.

The superconducting properties of SmFeAsO$_{0.85}$  and
NdFeAsO$_{0.85}$ were studied in the transverse-field (TF) $\mu$SR
experiments. The temperature scans were made after cooling the
samples from above $T_c$ down to 1.7~K in $\mu_0H=0.2$~T.
Following Hayano {\it et al.} \cite{Hayano79} it can be  shown
that the effect of fast fluctuations on the longitudinal and
transverse depolarization become similar. By taking this into
account the TF $\mu$SR data were analyzed by using the following
functional form:
\begin{equation}
P^{\rm TF}(t)=P^{\rm LF}(t)\exp\left[-\frac{(\sigma_{sc}^2+\sigma_{nm}^2)t^2}{2}\right]\cos(\gamma_\mu Bt+\phi).
 \label{eq:TF}
\end{equation}
Here $B$ is the average field inside the sample,  $\gamma_{\mu} =
2\pi\times135.5342$~MHz/T is the muon gyromagnetic ratio, $\phi$
is the initial phase, and $P^{\rm LF}(t)$ is described by
Eq.~(\ref{eq:ZF-LF}). $\sigma_{nm}$ denotes the muon-spin
relaxation rate caused by the nuclear moments and $\sigma_{sc}$ is
the additional component appearing below $T_c$ due to nonuniform
field distribution in the superconductor in the mixed state.
During the fit $\sigma_{nm}$ was fixed to the values obtained
above $T_c$.

\begin{figure}[htb]
\includegraphics[width=0.9\linewidth]{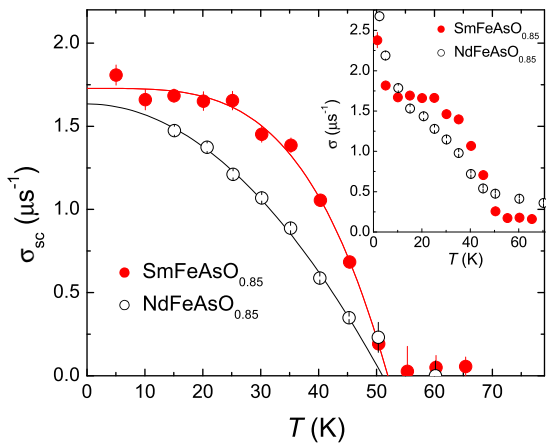}
\caption{(Color online) Temperature  dependences of
$\sigma_{sc}\propto\lambda^{-2}_{ab}$ for SmFeAsO$_{0.85}$ and
NdFeAsO$_{0.85}$ measured after field cooling the samples in
$\mu_0H=0.2$~T. The inset shows the total Gaussian depolarization
rate $\sigma=(\sigma_{sc}^2+\sigma_{nm}^2)^{1/2}$. }
 \label{fig:T-scan}
\end{figure}

Fig.~\ref{fig:T-scan} shows the $\sigma_{sc}(T)$ measured  in
$\mu_0H=0.2$~T. The inset shows the total Gaussian depolarization
rate $\sigma=(\sigma_{sc}^2+\sigma_{nm}^2)^{1/2}$. The fast
increase of $\sigma_{sc}$ below $T\sim$50~K is due to the
well-known fact that type-II superconductors exhibit a flux-line
lattice leading to spatial inhomogeneity of the magnetic
induction. As shown by Brandt \cite{Brandt88}, in a case of
anisotropic powder superconductor the second moment of this
inhomogeneous field distribution is related to the in-plane
magnetic penetration depth $\lambda_{ab}$ in terms of:
\begin{equation}
\langle \Delta
B^{2}\rangle=\frac{\sigma_{sc}^2}{\gamma_\mu^2}=0.0371\frac{\Phi_0^2}{\lambda_{eff}^{\
4}}= 0.0126\frac{\Phi_0^2}{\lambda_{ab}^{\ 4}},
 \label{eq:lambda_abs}
\end{equation}
where $\Phi_0=2.068\cdot10^{-15}$~Wb is the magnetic flux quantum.
Here we also take into account that in anisotropic superconductor
the effective magnetic penetration depth $\lambda_{\rm eff}$,
measured in $\mu$SR experiments, is solely determined by the in-plane penetration depth as $\lambda_{\rm eff}=1.31\lambda_{ab}$
\cite{Fesenko91}.

The data in Fig.~\ref{fig:T-scan} were fitted with the power  law
$\sigma_{sc}(T)/\sigma_{sc}(0)=1- (T/T_{c})^n$ with
$\sigma_{sc}(0)$, $n$, and $T_c$ as free parameters. Due to extra
Sm and, most probably, Nd orderings (see the inset in
Fig.~\ref{fig:T-scan}) only the points above 5~K for Sm and above
10~K for Nd substituted samples were considered.  The fit yields
$\sigma_{sc}(0)=1.73(5)$~$\mu$s$^{-1}$, $T_c=52.0(2)$~K,
$n=3.74(16)$ and $\sigma_{sc}(0)=1.63(5)$~$\mu$s$^{-1}$,
$T_c=51.0(3)$~K, $n=1.98(14)$ for SmFeAsO$_{0.85}$ and
NdFeAsO$_{0.85}$, respectively. From measured $\sigma_{sc}(0)$ the
absolute values of the in-plane magnetic penetration depth
obtained by means of Eq.~(\ref{eq:lambda_abs}) are
$\lambda_{ab}(0)=189(5)$~nm and $\lambda_{ab}(0)=195(5)$~nm for
SmFeAsO$_{0.85}$ and NdFeAsO$_{0.85}$. At the present stage we are
not going to discuss the temperature dependences of
$\sigma_{sc}(T)$ curves. We would only mention that the power law
exponent $n=3.74(16)$ is very close to the universal two-fluid
value $n\equiv4$, while $n=1.98(14)$ is close to $n=2$ which is
generally observed in a case of dirty $d-$wave superconductors.

\begin{figure}[htb]
\includegraphics[width=0.9\linewidth]{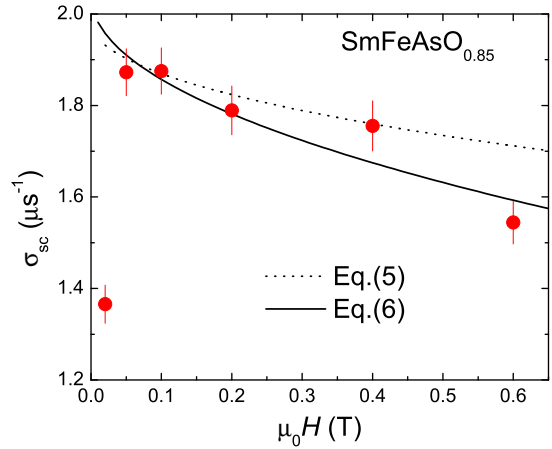}
\caption{(Color online) $\sigma_{sc}$ as a function of applied
field  $H$ for SmFeAsO$_{0.85}$. Each point was obtained after
field-cooling the sample from above $T_c$ to $T=15$~K in the
corresponding magnetic field. The solid and the dashed lines
represents the results of analysis by means of
Eqs.~(\ref{eq:sigma_vs_h}) and (\ref{eq:lambda_vs_H}). }
 \label{fig:Field-Scan}
\end{figure}

The magnetic field dependence of $\sigma_{sc}$ for
SmFeAsO$_{0.85}$  is shown in Fig.~\ref{fig:Field-Scan} (the
corresponding $\sigma_{nm}$ values were obtained from calibration
runs made at $T=65$~K). At low fields a maximum in
$\sigma_{sc}(H)$ is observed followed by a decrease of the
relaxation rate up to the highest fields.
Consideration of the ideal triangular vortex lattice of an
isotropic $s-$wave superconductor within the Ginsburg-Landau
approach leads to the following expression for the magnetic field
dependence of the second moment of the magnetic field distribution
\cite{Brandt03}:
\begin{eqnarray}
 \sigma_{sc}[\mu {\rm s}^{-1}]=4.83\times10^4
(1 - B/B_{c2}) \nonumber \\
 \left[1 + 1.21\left(1 - \sqrt{B/B_{c2}}\right)^3\right]&
\lambda^{-2}[{\rm nm}] .
 \label{eq:sigma_vs_h}
\end{eqnarray}
Here $B$ is the magnetic induction, which for applied field in the
region  $H_{c1}\ll H \ll H_{c2}$ is $B\simeq \mu_0 H$ ($H_{c1}$ is
the first critical field, and $B_{c2}=\mu_0H_{c2}$ is the upper
critical field). According to \cite{Brandt03},
Eq.~(\ref{eq:sigma_vs_h}) describes with less than 5\% error the
field variation of $\sigma_{sc}$ for an ideal triangular vortex
lattice and holds for type-II superconductors with the value of
the Ginzburg-Landau parameter $\kappa\geq5$ in the range of fields
$0.25/\kappa^{1.3}\lesssim B/B_{c2}\leq1$.
Since we are not aware of any reported values of the second
critical  field for SmFeAsO$_{0.85}$, we used
$B_{c2}(15$~K)$\sim80$~T obtained from $B_{c2}(0)\sim100$~T
reported by Senatore {\it et al.} \cite{Senatore08} for
SmO$_{0.85}$F$_{0.15}$FeAs. The black dotted line, derived by using
Eq.~(\ref{eq:sigma_vs_h}), corresponds to $\lambda=232$~nm.
Fig.~\ref{fig:Field-Scan} implies that the experimental
$\sigma_{sc}(H)$ depends stronger on the magnetic field than it is
expected in a case of fully gaped $s-$wave superconductor. As
shown by Amin {\it et al.} \cite{Amin00} the field dependent
correction to $\rho_s$ may arise from the nonlocal and nonlinear
response of a superconductor with nodes in the energy gap to the
applied magnetic field. The solid  line represent the result
of the fit by means of the relation:
\begin{equation}
\frac{\rho_s(H)}{\rho_s(H=0)}=
\frac{\sigma_{sc}(H)}{\sigma_{sc}(H=0)}= 1-K\cdot \sqrt{H},
 \label{eq:lambda_vs_H}
\end{equation}
which takes into account the nonlinear correction to $\rho_s$ for
a  superconductor with a $d-$wave energy gap  \cite{Vekhter99}.
Here the parameter $K$ depends on the strength of nonlinear
effect. Since Eq.~(\ref{eq:lambda_vs_H}) is valid for the
intermediate fields ($H_{c1}\ll H\ll H_{c2}$) only the points
above 0.02~T were considered in the analysis.

\begin{figure}[htb]
\includegraphics[width=0.8\linewidth]{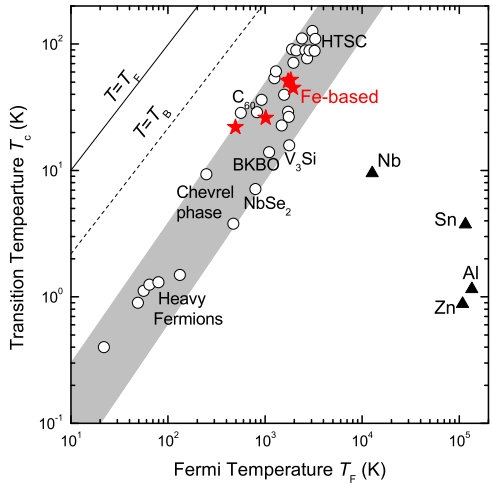}
\caption{(Color online) The superconducting transition temperature
$T_c$ {\it vs.} the effective Fermi temperature $T_F$. The
unconventional superconductors fall within a common band for which
$1/100<T_c/T_F<1/10$ as indicated by the grey region in the
figure. The dashed line correspond to the Bose-Einstein
condensation temperature $T_B$ (see Ref.~\onlinecite{Uemura91} for
details). The points for SmFeAsO$_{0.85}$ and  NdFeAsO$_{0.85}$
calculated from $\lambda_{ab}(0)$ values obtained in the present
study and that for LaO$_{1-x}$F$_x$FeAs ($x=0.1$, 0.075)
\cite{Luetkens08} and SmO$_{0.82}$F$_{0.18}$FeAs \cite{Drew08} are
shown by solid red stars.}
 \label{fig:Uemura}
\end{figure}

As a next step we focus on the Uemura classification scheme which
considers the correlation between the superconducting transition
temperature $T_c$ and the effective Fermi temperature $T_F$
determined from measurements of the penetration depth. Using this
parameterisation Uemura {\it et al.} \cite{Uemura91} confirm a
close correlation between $T_c$ and $T_F$. HTS's, heavy fermion,
organic, fullerene and Chevrel phase superconductors all follow a
similar linear trend with $1/100< T_c/T_F <1/10$, in contrast to
the conventional BCS superconductors (Nb, Sn, Al etc) for which
$T_c/T_F <1/1000$. The ''Uemura plot`` of $\log(T_c)$ {\it vs.}
$\log(T_F)$, shown in Fig.~\ref{fig:Uemura}, thus appears to
discriminate between the ''unconventional`` and ''conventional``
superconductors. The $T_B$ represents the Bose-Einstein
condensation temperature for a non interacting 3D Bose gas having
the boson density $n_B=n_s/2$ and mass $m_B=2m^\ast$.
Intriguingly, all the ''unconventional`` superconductors are found
to have values of $T_c/T_B$ in the range 1/3 to 1/30, thereby
emphasizing the proximity of these systems to Bose-Einstein like
condensation.

Following suggestions of Uemura {\it et al.} \cite{Uemura91}, the
effective Fermi  temperatures of the Fe-based superconductors were
calculated as:
\begin{equation}
k_BT_F=\hbar\pi  c_{int} \frac{n_s}{m^\ast}\propto c_{int}\sigma_{sc}.
 \label{eq:Tfermi}
\end{equation}
Here $n_s/m^\ast=\rho_s$ is the superfluid density, which within
the London approach  is proportional to $\lambda^{-2}$ and, thus,
to $\sigma_{sc}$
($n_s/m^\ast\propto\lambda^{-2}\propto\sigma_{sc}$), $n_s$ is the
charge carrier concentration, $m^\ast$ is the charge carrier mass,
and $c_{int}$ is the distance between the conducting planes.
The points for SmFeAsO$_{0.85}$, NdFeAsO$_{0.85}$ and that
obtained  from $\lambda_{ab}(0)$ values of LaO$_{1-x}$F$_x$FeAs
($x=0.1$, 0.075) \cite{Luetkens08} and SmO$_{0.82}$F$_{0.18}$FeAs
\cite{Drew08} are shown in Fig.~\ref{fig:Uemura} by solid red
stars. As is seen the Fe-based superconductors follow the same
linear trend as is established for various unconventional
materials suggesting that they all probably share the common
condensation mechanism.


To conclude, measurements of the in-plane magnetic field
penetration  depth $\lambda_{ab}$ in superconductors
SmFeAsO$_{0.85}$ ($T_c\simeq52$~K) and NdFeAsO$_{0.85}$
($T_c\simeq51$~K) were carried out by means of muon-spin-rotation.
The absolute values of $\lambda_{ab}$ at $T=0$ were estimated to
be 189(5)~nm and 195(5)~nm for Sm and Nd substituted samples,
respectively. The analysis within the Uemura classification
scheme, considering the correlation between the superconducting
transition temperature $T_c$ and the effective Fermi temperature
$T_F$ reveal that  both families of Fe-based superconductors (with
and without fluorine) falls to the same class of unconventional
superconductors.
%


This work was performed at the Swiss Muon Source (S$\mu$S),  Paul
Scherrer Institute (PSI, Switzerland).

\end{document}